\documentclass[pdflatex,sn-mathphys,twocolumn]{sn-jnl}
\usepackage{aas_macros}

\usepackage{anyfontsize}
\usepackage{natbib}

\jyear{2024}
\renewcommand{\thesubsection}{\arabic{subsection}}

\usepackage{newfloat}
\DeclareFloatingEnvironment[name={Supplementary Fig.}]{suppfigure}

\usepackage{xr}
\begin{document}
\title[$~$]{\vspace{-2.5cm} Gravitational Lensing Reveals Cool Gas within 10-20 kpc around a Quiescent Galaxy}

\author*[1,2,3]{\fnm{Tania M. }\sur{Barone}}\email{tbarone@swin.edu.au}
\author[1,2]{\fnm{Glenn G. }\sur{Kacprzak}}
\author[4]{\fnm{James W. }\sur{Nightingale}}
\author[1,2,5]{\fnm{Nikole M. }\sur{Nielsen}}
\author[1,2]{\fnm{Karl }\sur{Glazebrook}}
\author[2,3,6]{\fnm{Kim-Vy H. }\sur{Tran}}
\author[7]{\fnm{Tucker }\sur{Jones}}
\author[1,2]{\fnm{Hasti }\sur{Nateghi}}
\author[7]{\fnm{Keerthi }\sur{Vasan G. C.}}
\author[2,3]{\fnm{Nandini }\sur{Sahu}}
\author[1,2]{\fnm{Themiya }\sur{Nanayakkara}}
\author[8]{\fnm{Hannah }\sur{Skobe}}
\author[9]{\fnm{Jesse }\sur{van de Sande}}
\author[10]{\fnm{Sebastian }\sur{Lopez}}
\author[9]{\fnm{Geraint F. }\sur{Lewis}}

\affil[1]{\orgdiv{Centre for Astrophysics and Supercomputing}, \orgname{Swinburne University of Technology},\orgaddress{\city{Melbourne}, \country{Australia}}}

\affil[2]{\orgname{ARC Centre for Excellence in All-Sky Astrophysics in 3D (ASTRO 3D)}, \orgaddress{\city{Canberra}, \country{Australia}}}

\affil[3]{\orgdiv{School of Physics}, \orgname{University of New South Wales}, \orgaddress{\city{Sydney}, \country{Australia}}}

\affil[4]{\orgdiv{Centre for Extragalactic Astronomy, Department of Physics}, \orgname{Durham University}, \orgaddress{\city{Durham}, \country{United Kingdom}}}

\affil[5]{\orgdiv{Homer L. Dodge Department of Physics and Astronomy}, \orgname{The University of Oklahoma}, \orgaddress{Norman}, \country{United States of America}}

\affil[6]{\orgdiv{Center for Astrophysics}, \orgname{Harvard \& Smithsonian}, \orgaddress{\city{Cambridge}, \state{United States of America}}}

\affil[7]{\orgdiv{Department of Physics and Astronomy}, \orgname{University of California Davis}, \orgaddress{\city{Davis}, \country{United States of America}}}

\affil[8]{\orgdiv{Department of Astronomy and Astrophysics}, \orgname{University of Chicago}, \orgaddress{\city{Chicago}, \country{United States of America}}}

\affil[9]{\orgdiv{Sydney Institute for Astronomy}, \orgname{The University of Sydney}, \orgaddress{\city{Sydney}, \country{Australia}}}

\affil[10]{\orgdiv{Departamento de Astronomía}, \orgname{Universidad de Chile}, \orgaddress{\city{Santiago}, \country{Chile}}}

\abstract{While quiescent galaxies have comparable amounts of cool gas in their outer circumgalactic medium (CGM) compared to star-forming galaxies, they have significantly less interstellar gas. However, open questions remain on the processes causing galaxies to stop forming stars and stay quiescent . Theories suggest dynamical interactions with the hot corona  prevent cool gas from reaching the galaxy, therefore predicting the inner regions of quiescent galaxy CGMs are devoid of cool gas. However, there is a lack of understanding of the inner regions of CGMs due to the lack of spatial information in quasar-sightline methods. We present integral-field spectroscopy probing 10--20~kpc (2.4--4.8 R\textsubscript{e}) around a massive quiescent galaxy using a gravitationally lensed star-forming galaxy. We detect absorption from Magnesium (MgII) implying large amounts of cool atomic gas (10\textsuperscript{8.4} -- 10\textsuperscript{9.3} M\textsubscript{$\odot$} with T$\sim$10\textsuperscript{4} Kelvin), in comparable amounts to star-forming galaxies. Lens modeling of Hubble imaging also reveals a diffuse asymmetric component of significant mass consistent with the spatial extent of the MgII absorption, and offset from the galaxy light profile. This study demonstrates the power of galaxy-scale gravitational lenses to not only probe the gas around galaxies, but to also independently probe the mass of the CGM due to it's gravitational effect.}

\keywords{Quiescent Galaxies, Circumgalactic Medium, Gravitational Lensing, Lens Modelling}

\maketitle
\small

\begin{figure*}
    \centering
    \includegraphics[width=1.0\textwidth]{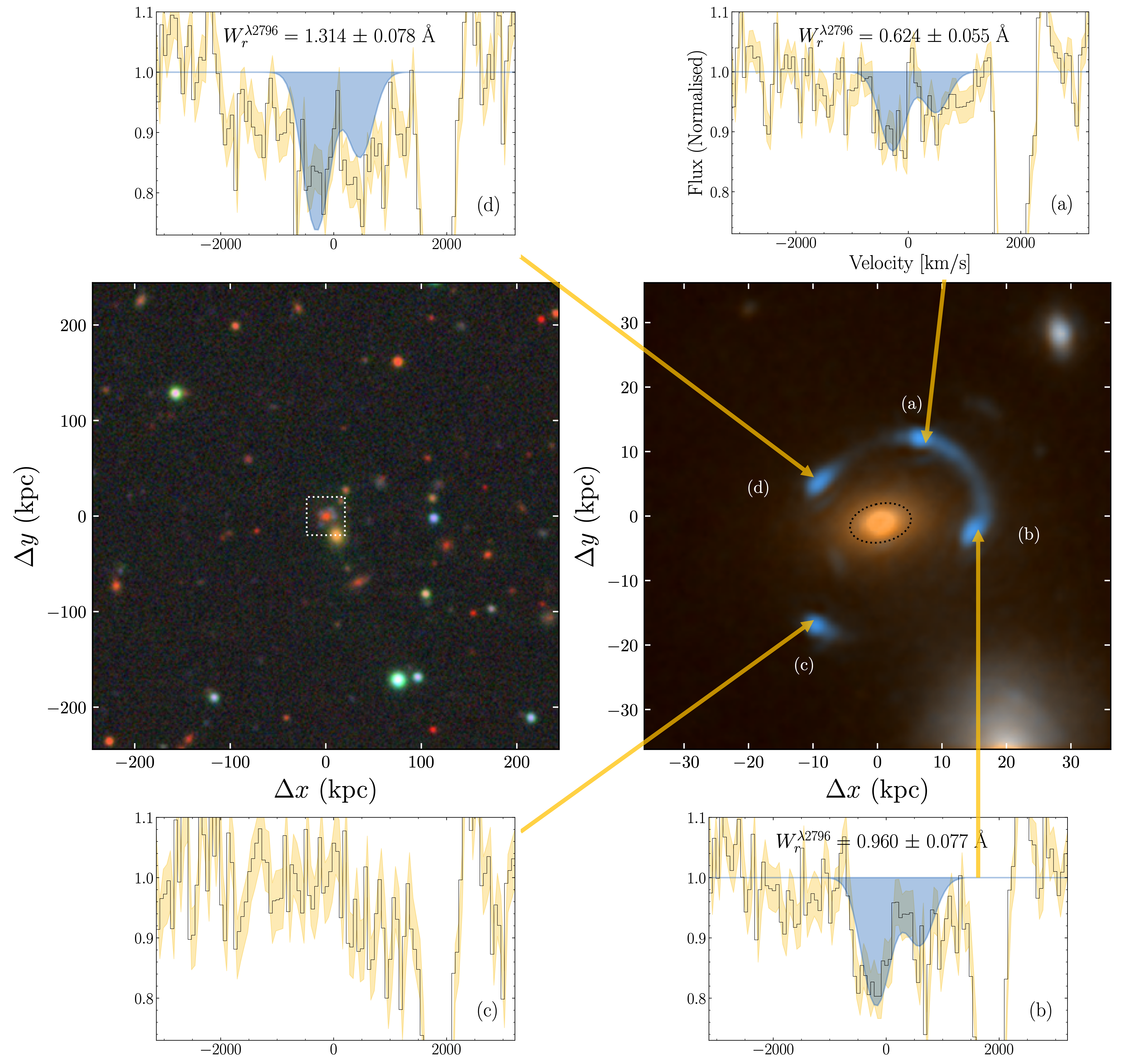}
    \caption{\textbf{Map of MgII absorption in the 4 arcs of the gravitational lens.} \textit{Left}: DECaLS $2'\times2'$ image demonstrating the relatively isolated local environment. The white box shows the field of view of the right panel. \textit{Right}: Combined HST/WFC3 F606W and F140W image of J0755 with labels for the four bright arcs of the Einstein cross. The black dotted line encapsulates the galaxy major axis ratio and $1 R_\mathrm{e}$ \cite{Ritondale2019a}. \textit{(a - d)}: Voigt profile fits to MgII in the KCWI spectra of the 4 arcs. The black line shows the data, the yellow lines show $\pm 1\sigma$ uncertainty, and the blue lines show the fitted Voigt profiles. The rest-frame equivalent width of $\lambda 2796$ absorption, $W^{\lambda 2796}_{r}$, is listed at the top. The strong absorption at 1770 km s$^{-1}$ (4845 {\AA} observed wavelength) is interstellar CII from the arc.}
    \label{fig:hubble_kcwi_img}
\end{figure*}

\section*{Introduction}

Why galaxies remain quiescent despite the presence of gas in their halos remains a puzzle in galaxy evolution models. In particular, the discovery of cool (T$\sim 10^4$) gas situated within the hot ($T\sim 10^{6-7}$) corona \citep{Thom2012} of massive passive galaxies contradicts the previously held thesis that these galaxies quenched due to a lack of cool gas accreting from their circumgalactic mediums (CGMs) \cite[][]{Stewart2011a}. While recent observations have firmly established the prescence of cool gas around massive quiescent galaxies out to 500 kpc \citep{Gauthier2010,Huang_Yun-Hsin2016}, state-of-the-art models struggle to accurately reproduce the diversity of observed properties \citep{Nelson2020}. Furthermore, it remains unclear whether the cool gas present at large impact parameters is able to survive infall and accrete onto the galaxy, or whether the clouds are destroyed via interactions with the surrounding hot medium \citep{Afruni2019}. As a result, the innermost regions ($\sim 0 - 30$ kpc) of the CGM are of particular interest for quenching models.\\

Observations of the inner CGM around quiescent galaxies have primarily relied on quasar sightlines. However quasar sightlines are limited by a lack of spatial information on the level of individual galaxies due to quasars providing only a needle beam sampling of the CGM in question. Additionally, quasar sightlines studies are unable to control for the impact parameter and viewing angle probed, both of which strongly influence CGM properties \citep[e.g. ][]{Werk2013,Nielsen2013}. While multiply lensed quasars mitigate some of these issues by providing multiple probes at different viewing angles \cite{Chen_Hsiao-Wen2014,Zahedy2016,Zahedy2020}, each lensed image still only provides a needle-like probe, and sample sizes are limited by their rarity. The small spatial area probed by quasars is a critical issue, as simulations indicate the cool CGM to be clumpy and non-isotropic \cite{Peeples2019,Peroux2020}. Given the typical size of the broad line region of an AGN ($\sim 10^{-2}$pc \cite{Mandal2021}) and the typical size of a massive elliptical galaxy's halo ($R_{vir} \sim 500$ kpc) a quasar probes $\sim 10^{-3}$ pc$^2$ of a medium spanning a cross-sectional area of $10^{12}$pc$^2$. It is therefore extremely difficult to accurately characterise the total gas present around galaxies with quasar sightlines alone, as they are highly vulnerable to the stochastic nature of the CGM distribution. As a result, measuring properties such as the gas covering fraction requires stacking of large samples. However, such samples cannot fully control for all properties know to influence the CGM (e.g. azimuthal angle \cite{Bordoloi2011,Bouche2012,Kacprzak2012}, impact parameter
\cite{Steidel1994,Kacprzak2008,Nielsen2013}, stellar mass \cite{Bordoloi2018,Smailagic2023}, star formation rate \cite{Tumlinson2011}, and environment \cite{Chen2010,Bordoloi2011,de_Blok2018,Nielsen2018,Dutta2020}).\\

Alternatively, gravitationally lensed background galaxies allow us to probe foreground targets with exquisite spatial detail. A typical gravitational arc can probe an area of $\sim$100 kpc$^2$ providing $10^{11}$ times more cross-sectional area than a quasar (see Methods subsection Arc Areas and Impact Parameters). This significantly larger spatial coverage leads to more representative measurements on the chemical abundances, temperature and density of the gas. Furthermore, due to the significantly larger coverage, gravitational lenses can measure the covering fraction of the gas directly around a single galaxy. The first studies to employ this arc-tomographic approach \cite{Lopez2018,Lopez2020,Mortensen2021,Tejos2021,Fernandez-Figueroa2022} used cluster scale lenses to probe the CGM of intervening star-forming galaxies. However, galaxy-galaxy strong lenses consisting of a massive quiescent deflector and a highly star-forming background source are significantly more abundant than cluster-scale lenses.\\

Over the next decade of order $10^6$ galaxy-galaxy strong lenses will be discovered by cosmological surveys such as Euclid and the Legacy Survey of Space and Time \cite{Collett2015}, a three orders of magnitude increase over the hundreds of systems that are currently known \citep{Bolton2008,Sonnenfeld2013,Bolton2012,Shu2016a,Jacobs2019a,X_Huang2020,X_Huang2021,Tran2022}. This is also two orders of magnitude above the number of lensed quasars that will be found ($\sim 10^3$, \cite{Yue2022}). With galaxy-scale lensing preferentially occurring around massive quiescent galaxies, we show for the first time how these systems can be used to characterise the innermost regions of the circumgalactic medium around quiescent galaxies. Furthermore, due to the simple lens mass profile, we show for the first time that the mass of the CGM itself is sufficient to influence gravitational lens models.\\

We present a study of the gravitational lens SDSSJ075523.52+344539.5 (hereafter J0755) comprised of a single quiescent deflector at redshift $z=0.7221 \pm 0.0002$ \cite{Dawson2013} and a star-forming source at $z=2.6347$ \cite{Shu2016a} quadruplely lensed into an Einstein cross (Fig. \ref{fig:hubble_kcwi_img}). The 4 arcs range from 11 to 19 kpc from the deflector in projected distance and probe cross-section areas between $13.6$ and $39.2$ kpc$^2$ (Table \ref{tbl:fits_summary}). The deflector has a high stellar mass $\log M_*/M_\odot = 10.79^{+0.07}_{-0.01}$ and high total mass with $\log M_{\rm E}/M_\odot = 12.13$, where $M_{\rm E}$ is the total mass enclosed by the lensing radius (Methods subsection Lens Model). The deflector contains an old stellar population with a mean luminosity-weighted age of 3.3 Gyr and a star formation rate (SFR) of $0.49^{+0.36}_{-0.38} \;M_\odot/$yr (Methods subsection Stellar Properties). The high mass and low SFR are further supported by the deflector's high stellar velocity dispersion of $304 \pm 13$ km s$^{-1}$ (Methods subsection Stellar Properties) and no detected nebular emission lines in the observed spectrum. J0755 lives in a relatively isolated environment (Fig. 1 and Methods subsection Environment) so we can rule out the possibility of close neighbours influencing the results.\\

\section*{Results}

To search for cool gas absorption signatures in the brightly lensed arcs, we observed J0755 with the Keck Cosmic Web Imager (KCWI) \citep{Morrissey2018} at the Keck Observatory. The absorption from the MgII doublet at $\lambda=2796, 2803${\AA} reveals a significant amount of cool gas along the line-of-sight toward three of the four arcs. Fig. \ref{fig:hubble_kcwi_img} shows the MgII absorption in the 4 different positions of the Einstein cross labelled (a) to (d). Of the four arcs, arc (b) shows the clearest MgII absorption, and we also find MgII in arcs (a) and (d) (Table \ref{tbl:fits_summary}). Despite the similar impact parameters of arcs (a), (b), and (d), there is azimuthal variation in the absorption strength of MgII.\\ 

\begin{table*}
\resizebox{\textwidth}{!}{
\begin{tabular}{ccccccccc}
\hline
Arc & $\rho$ & Area & $W^{\lambda 2796}_r$ & $\log_{10} N_{\rm Mg_{II}}$ & b & $\log_{10} N_{\rm H_I}$ & $\Delta v$ & $\log_{10} M_{\rm H_I}$ \\
 & (kpc) & (sq") & ({\AA}) & (atoms cm$^{-2}$) & (km/s) & (atoms cm$^{-2}$) & (km/s) & $ (M_\odot)$ \\
(1) & (2) & (3) & (4) & (5) & (6)& (7) & (8) & (9)\\\hline\hline
a & 14.4 & 0.52 & 0.62 $\pm$ 0.06 & 13.24 +/- 0.05 & 297 $\pm$ 47 & 19.1 - 20.3 & $-227 \pm 53$ & 6.2 - 7.4 \\
b & 14.6 & 0.74 & 0.96 $\pm$ 0.08 & 13.51 +/- 0.04 & 332 $\pm$ 44 & 19.5 - 20.4 & $-126 \pm 51$ & 6.8 - 7.7 \\
c & 18.9 & 0.26 & $<$ 0.315 & $<$ 13.2 &  &  &  &  \\
d & 11.1 & 0.75 & 1.31 $\pm$ 0.08 & 13.58 +/- 0.05 & 305* & 19.7 - 20.4 & $-255 \pm 55$ & 7.1 - 7.8 \\
\hline
\end{tabular}
}\caption{\textbf{Summary of CGM parameters.} Column (1) arc ID, (2) impact parameter, (3) area probed by the arc, (4) rest-frame equivalent width of MgII $\lambda 2796$, (5) MgII column density, (6) velocity dispersion, (7) HI column density converted from $W_r^{\lambda 2796}$\cite{Menard_Chelouche2009}, (8) velocity offset with respect to the galaxy, (9) gas mass in HI probed by the arcs \cite{Werk2013}, * The velocity width of arc (d) was unconstrained by the data (see Methods section KCWI Observations) so the $b$ parameter was fixed during the fit.}\label{tbl:fits_summary}
\end{table*}

We estimate the HI gas mass directly probed by arcs (a), (b), and (d) to be between $10^{6.2} - 10^{7.4}$, $10^{6.8} - 10^{7.7}$, and $10^{7.1} - 10^{7.8}M_\odot$ respectively (Methods subsection KCWI Observations). If we assume the three detections are representative of the total HI present within 18 kpc, i.e. that the gas is uniformly distributed, we estimate between $10^{8.4} - 10^{9.3} M_\odot$ of HI gas. Compared to the stellar mass of the galaxy ($M_* = 10^{10.79} M_\odot$), these HI masses are between $0.4 - 3\%$ of the galaxy's stellar mass, which is consistent with the $M_{HI}/M_*$ fraction found within the stellar body and inner $\sim10$'s of kpc of $\sim20 - 25\%$ of similar mass early-type galaxies at $z=0$ \cite{Serra2012} (see Supplementary Fig. \ref{fig:atlas3d_HI_gas_mass}).\\

The column densities and gas masses presented in Table \ref{tbl:fits_summary} are based on an analysis assuming unity coverage of the gas over the arcs. Arcs (a) and (b) are well fit by the unity covering fraction model and so if we instead allow for partial covering in the Voigt profiles we find that partial covering models only provide a marginal improvement in goodness of fit. Conversely, arc (d) is likely to be only partially covered based on the doublet ratio of the $\lambda \lambda 2796, 2803$ absorption features (which is 2:1 in the case of unity coverage and unsaturated lines), and there is a significant improvement in the goodness of fit when allowing for partial covering. Importantly, partial covering significantly increases the column density estimates: $\log N_{MgII}$ in arcs (a) and (b) can increase by nearly an order of magnitude, and the most severely affected, arc (d), $\log N_{MgII}$ can increase by 2-3 orders of magnitude. Naturally, larger column densities lead to larger gas mass estimates, and therefore possible partial covering would only increase our HI mass estimates. Therefore we present here the results from the unity covering fraction model, and consider that the column densities and gas masses are lower limits in the case of partial covering.\\

It is interesting to note however that the data does imply partial coverage. Individual clouds range on scales of 0.1–30 kpc \citep{Stocke2013} and are therefore much larger than a quasar. As a result, quasars do not show partial coverage, and therefore the distribution of clouds around galaxies can only be inferred from stacked quasar sightline samples and comparing the frequency of incidence of the sightlines with a cloud (the fractional coverage). On the other hand, using gravitational lenses we can statistically measure the distribution (size and coverage) of clouds around a single galaxy, without the need for stacking large samples. Specifically, we find partial coverage fractions of 0.81 for arcs (a) and (b) and 0.86 for arc (d) minimise the reduced $\chi^2$ of the fits, indicating the medium probed is highly clumped. Furthermore, the difference in partial coverage between arc (d) and arcs (a) and (b) suggests azimuthal variation in the density of cool clouds at $\sim 10-15$kpc.\\

\begin{figure*}
    \centering
    \includegraphics[width=\textwidth]{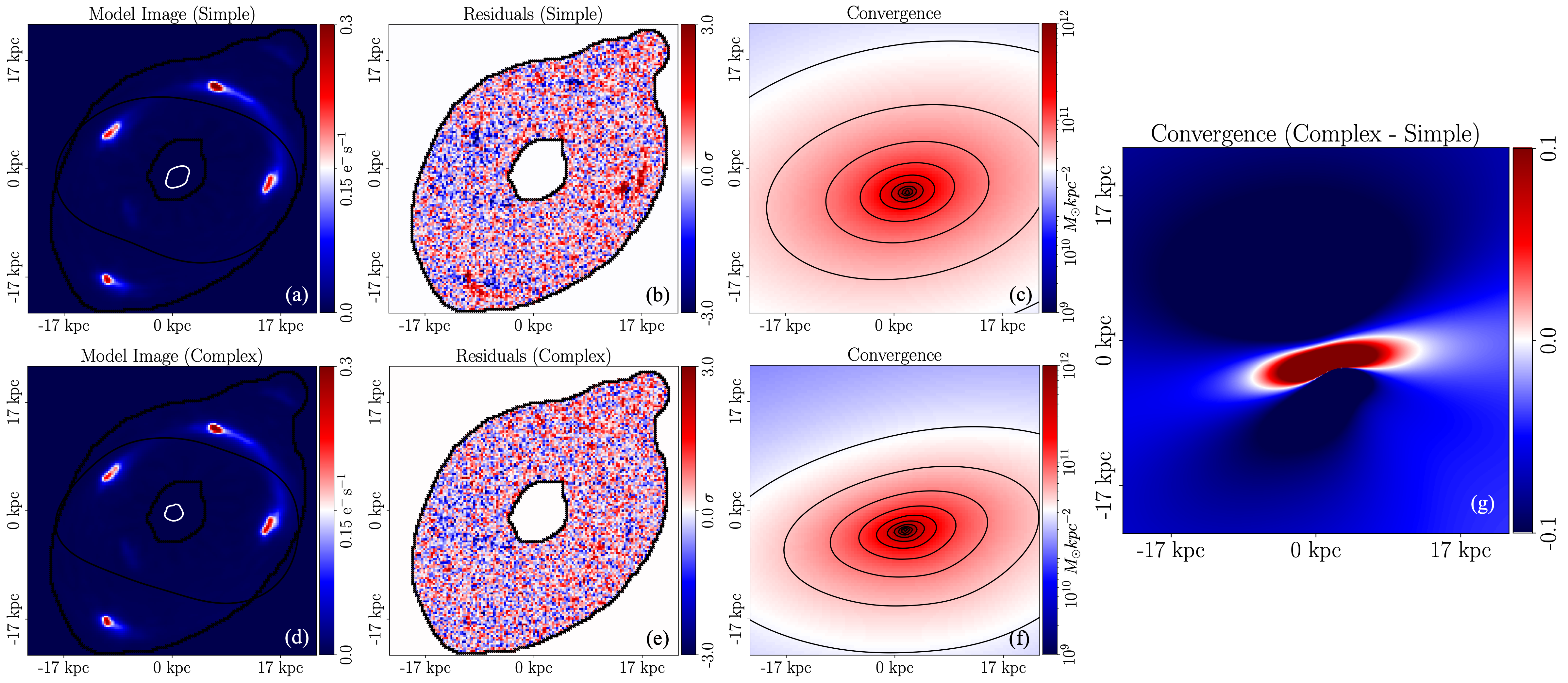}
    \caption{\textbf{Lens models for J0755 in the HST/F606W filter}. The top row shows the power law (simple) model, and the bottom row shows the power law plus multipoles (complex) model. \textit{(a) and (d)}: Model of the lensed source. \textit{(b) and (e)} Residuals from the model (model - lens subtracted data). \textit{(c) and (f)}: Convergence of the maximum likelihood mass model. \textit{g}: Difference between the convergence maps for the simple and complex model. The more complex mass model significantly improves the residuals of the fit. The contours overlaid on the right panels show that it does this by asymmetrically squeezing the mass distribution towards arcs b and d where the strongest KCWI detections are found.}
    
    \label{fig:lens_model}
\end{figure*}

J0755 is known as a difficult-to-model lens, despite it having a simple geometry \cite{Shu2016b,Ritondale2019a,Ritondale2019PhDT,Nightingale2022}. Previous studies suggested and investigated numerous possible explanations, including a complicated mass profile in the deflector and local environment \cite{Shu2016b,Ritondale2019b}, or line of sight dark matter subhalos \cite{Ritondale2019b,Nightingale2022}. Given our results showing significant amounts of cool gas intersected by 3 of the 4 arcs, this raises the question of whether the mass of the CGM could be significantly contributing to how light from the background source is being deflected. If this CGM gas is asymmetrically distributed around the galaxy, it could have a measurable impact on the lensing and would explain the inability of previous models to precisely reproduce the lensing light profile.\\

Fig. \ref{fig:lens_model} shows the results of lens modeling of the F606W image, (Methods subsection Lens Model). The top row shows fits using an elliptical power-law (PL) mass distribution with external shear, which represents the summed mass of stars and dark matter. The PL is commonly used to model strong lenses and typically fits lens imaging data to the noise level (e.g. \cite{Etherington2023}). However, Fig. \ref{fig:lens_model} shows this model leaves significant residuals in all four arcs.\\

To include the lensing contribution of the diffuse, asymmetric gas component, we simultaneously add first, second, third and fourth order multipole expansions to the PL \citep{Chu2013}. Multipoles add azimuthal freedom to the mass distribution allowing it to capture complexities like an offset centre of mass and radial twists. The bottom row of Fig. \ref{fig:lens_model} shows fits for the PL with multipoles, where the significant residuals seen in the PL model are removed and the data is successfully fitted. The PL with multipoles model is favoured over the PL with a Bayes factor of 574 (corresponding to an over $10\sigma$ result) providing definitive evidence that the PL with multipoles model is capturing genuine mass structure.\\

Panels (c) and (f) of Fig. \ref{fig:lens_model} show the convergence maps of the PL and PL with multipoles models. The PL shows elliptically symmetric iso-density contours, whereas the PL with multipoles has a distinct asymmetry, with squeezing inwards in the direction of arcs (b) and (d), which exhibit the strongest MgII absorption. This is more clearly seen in panel (g), which presents the difference between the convergence maps of the two models. This asymmetry is driven primarily by the first order multipole, which has a magnitude of $\sim 12\%$, indicating a mass structure that is significantly offset from the stars and dark matter. A diffuse but significant amount of gas, offset from the stars, inside the Einstein radius is the most plausible explanation for the asymmetric mass distribution inferred via lensing.\\

Lens modeling of J0755 suggests it is a unique system, however expectations from quasar sightline studies show that a cool CGM around quiescent galaxies is not uncommon, at least at large impact parameters \cite{Thom2012,Huang_Yun-Hsin2016}. We quantify how unique it is from a lensing perspective by fitting 15 other strong lenses from the BELLS-GALLERY sample studied in \cite{Nightingale2022} with the PL plus multipoles model. We find four systems where Bayesian model comparison favours the PL with multipoles model, and in all four lens models the first order multipole exceeds $5\%$, indicating an offset mass component. The remaining 11 systems are typically lower signal-to-noise, therefore model comparison disfavouring the PL with multipoles does not necessary indicate it is not a more accurate description of their mass, but rather better data is necessary to infer this. Therefore, these results have important implications for strong lens modelling in general.\\

\begin{figure*}
    \centering
    \includegraphics[width=1.0\textwidth]{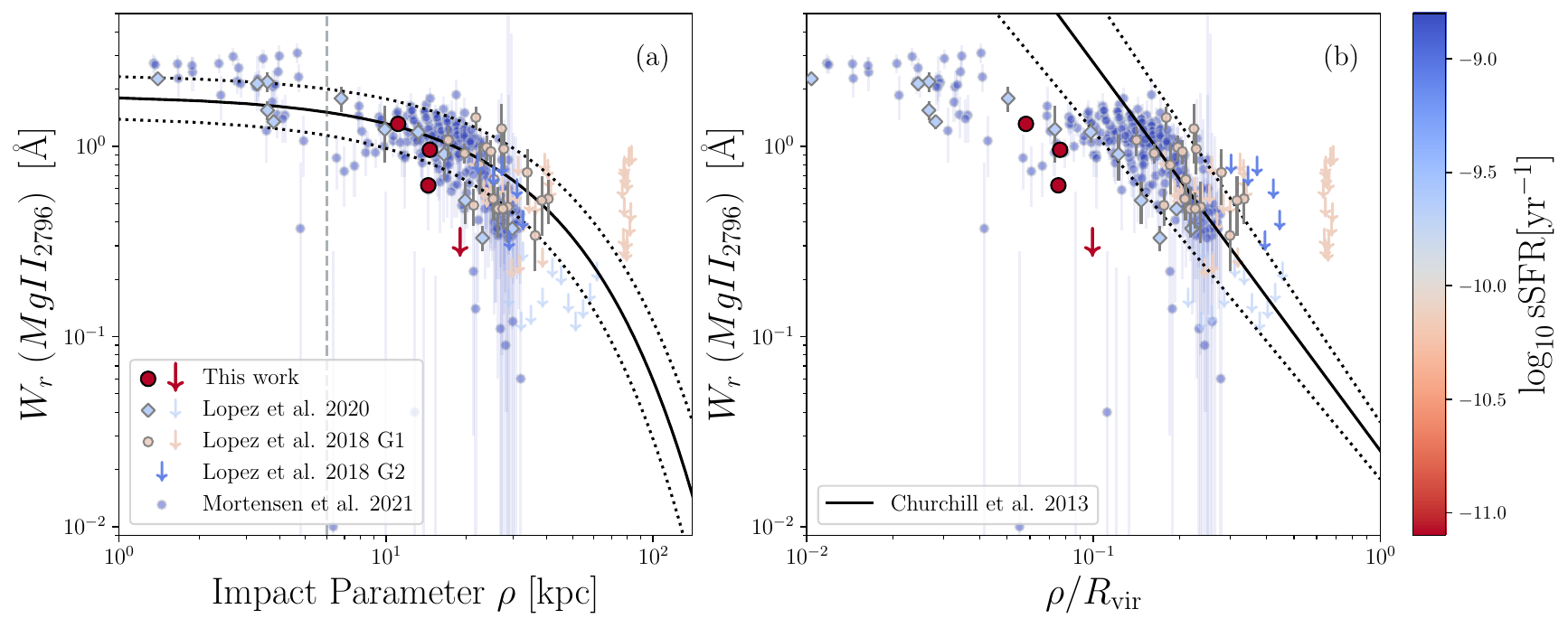}
    \caption{\textbf{Comparison of J0755 to star-forming galaxies probed with cluster scale arcs.} \textit{(a)}: Equivalent width of MgII versus impact parameter, with the relation expected for star-forming galaxies \cite{Nielsen2013}. The grey dashed vertical line is the expected boundary between interstellar and circumgalactic gas from $z\sim 0.1$ star-forming galaxies \cite{Kacprzak2013}. \textit{(b)}: Equivalent width of MgII versus impact parameter normalised by the virial radius, with the relation expected for star-forming galaxies \cite{Churchill2013}. The vertical bars show the $1\sigma$ uncertainty for each point, while the down arrows represent upper limits.}
    
    \label{fig:literature_comparison}
\end{figure*}

\section*{Discussion}

Massive passive galaxies reside in halos containing diffuse, metal-enriched, multi-phase gas extending out to $\leq 500$kpc \cite{Huang_Yun-Hsin2016} as revealed from absorption signatures imprinted by the gas on spectra of background quasars. However, quasar sightlines lack spatial information and typically probe the CGM at large galaxy--quasar separations, limiting our ability to characterise the clumpy and non-isotropic gas distribution \cite{Peeples2019,Peroux2020}. As a result, quasar absorption line studies of the CGM are unable to constrain total gas masses in individual galaxies and must heavily rely on statistically built galaxy halos to assume any constrains on the presence of gas and its mass. Conversely, using gravitational lenses provides a direct representative measurement of the CGM clouds in that region, as opposed to a probabilistic incidence of clouds along a needle sightline, and thus provide a greater challenge for simulations to reproduce.\\

In Fig. \ref{fig:literature_comparison} we show how the equivalent widths of MgII ($W_r^{\lambda 2796}$) measured here compare to other systems studied using a similar spatially resolved gravitational lensing technique using cluster scale arcs. All three studies \cite{Lopez2018}, \cite{Lopez2020} and \cite{Mortensen2021} probe the CGM of lower mass star-forming systems at similar redshifts as J0755 ($z = 0.734 - 0.98$ cf. $z_{\rm J0755} = 0.7224$). However, despite the systems probed by \cite{Mortensen2021}, \cite{Lopez2018} and \cite{Lopez2020} being significantly different in both stellar mass and star-formation rate, they have similar amounts of cool gas in their inner CGM compared to J0755. The stellar masses of these systems range from $M_* = 10^{9.0 - 9.8} M_\odot$ and have $\log \rm sSFR/yr^{-1} =  -10.15$ to $-8.8$. By comparison, J0755 is approximately an order of magnitude more massive ($M_* = 10^{10.79} M_\odot$) and significantly less star-forming with $\log \rm sSFR/yr^{-1}=-11.0$.\\

The vertical dotted line in panel (a) of Fig. \ref{fig:literature_comparison} shows the expected boundary between the ISM and CGM for $z \sim 0.1$ star-forming systems \cite{Kacprzak2013}. While the literature values with $\rho < 6$~kpc are slightly above the relation \cite{Nielsen2013} since the MgII absorption likely arises form both the ISM and CGM, the red points (J0755) are fully consistent with originating from CGM gas. The $W_r^{\lambda 2796}$ values from J0755 are consistent with the distributions from star-forming galaxies (most notably with the \cite{Mortensen2021} detections which also occur at small impact parameters). All 3 detections (and the upper limit for arc c) are consistent within 2$\sigma$ of the relation for star-forming systems \cite{Nielsen2013}. Scaling the impact parameter $\rho$ by the virial radius (panel b), there is still no significant difference between J0755 and the star-forming systems despite their significantly different virial radii (for J0755 $R_{\rm vir} = 191$ kpc compared to $73-135$ kpc for the star-forming systems). \\ 

In Supplementary Fig. \ref{fig:quasar_literature_comparison} we further compare J0755 to equivalent widths measured for quiescent (panel a) and star-forming (panel b) galaxies probed using quasar sightlines. Note however that given the difference in beam size between a quasar and a lensed galaxy arc, the measurements for J0755 are a result of a culmination of clouds within a large area compared to a needle beam intercepting just a few clouds in the literature studies.\\

The median velocities of the absorption in arcs (a) and (b) are blueshifted with respect to the deflector (227 and 126 km s$^{-1}$ respectively) but within expectations for gravitationally bound gas in a halo of this mass, which has an escape velocity of $\sim 600$ km s$^{-1}$ assuming the stellar-to-halo mass relation of \cite{Girelli2020} and an NFW halo profile \citep[see also][]{Huang_Yun-Hsin2016}. The absorption profiles are relatively broad ($297 \pm 47$, $332 \pm 44$ km s$^{-1}$) and consistent with the stellar velocity dispersion of the galaxy ($304 \pm 13$ km s$^{-1}$). Notably, the profiles are significantly broader than the absorption profiles measured for the lower mass star-forming galaxy of \cite{Mortensen2021} ($<45$ km $s^{-1}$, see \cite{Tejos2021}), and it is therefore unlikely the absorption is originating from an undetected low-surface brightness neighbour. We note however, that the gas probed is likely comprised of many smaller clouds moving at a range of velocities and containing lower internal dispersions, and so it is unlikely that any individual cloud has dispersions as high as $\sim 300$ km s$^{-1}$ \cite{Peeples2019}. Although, if the gas clouds are too dynamically hot to successfully accrete onto the galaxy, this provides a simple solution as to why the galaxy has remained quiescent despite this large cool gas reservoir at close impact parameters.\\

We can use the kinematic properties of the absorption profiles to consider whether this gas is likely to be inflowing or outflowing. For typical star-forming disk galaxies, accreting gas tends to co-rotate with the galaxy around the major axis at velocities less than or similar to the rotation velocity of the galaxy \cite{Steidel2002,Kacprzak2010,Ho2017}. However, given this galaxy is a massive passive galaxy, it may not have organised rotation driven by cosmic filaments and the fact that all the detections are blueshifted may instead be a sign that accretion is isotropic. Alternatively, gas accreted through wet mergers may display disturbed kinematics, and could lead to the blueshifted kinematics observed. The deflector is too faint at the wavelengths probed by Keck/KCWI to constrain its spatially-resolved stellar kinematics, however if the $304 \pm 13$ km s$^{-1}$ velocity dispersion is driven by rotation, then the MgII detections have a comparable velocity offset. Although, without knowing the direction of rotation from spatially resolved kinematics it's impossible to say for certain whether the absorber kinematics are indicative of accretion.\\

Considering the virial mass of the galaxy ($M_{\rm halo} = 10^{12.2} M_\odot$ \cite{Girelli2020}) is above the expected critical threshold mass at which galaxies switch from cold-mode to shock induced hot-mode accretion ($M_{sh} \sim 10^{12} M_\odot$) \cite{Dekel_Birnboim2006,Stewart2011b}. Therefore radiatively cooled clouds accreted from the hot halo are unlikely to survive to the inner halo regions (\cite{Binney2009,Nipoti2010,Afruni2023}; although the survival rate of cool clouds seeded from hot halos is not fully understood, e.g. \cite{Voit2017,Sobacchi_Sormani2019,Stern2024}). We therefore consider whether the existence of such a large cool gas reservoir in this regime could originate instead from outflows. Assuming then the gas originated within the galaxy and was ejected in star-formation driven outflows, we can use the luminosity-weighted stellar population age, 3.3 Gyr (Methods subsection Stellar Properties), as a proxy for the time since the galaxy stopped actively forming stars. Assuming the gas was ejected 3.3 Gyr ago, to reach distances of $\geq 11.1 - 14.6$ kpc it must therefore have had to travel at $\geq 4$ km s$^{-1}$. However, outflows tend to have significantly higher velocities of $150 - 200$ km s$^{-1}$ \cite{Bordoloi2014} which would therefore imply a much more recent period of star-formation (within the last 100 Myr) which is inconsistent with the deflector's rest-frame optical spectrum. \\

Another possibility is that the gas originated from an AGN-driven outflow. The galaxy currently does not host an AGN as evidenced by the lack of emission in the rest-frame optical spectrum. Specifically, no [OII] is detected in the OSIRIS spectrum (shown in Supplementary Fig. \ref{fig:deflector_OSIRIS}), and no [OIII] or $H_\beta$ is detected in the SDSS eBOSS spectrum \citep{Dawson2013}. Considering an AGN-driven outflow travelling at $1000$ km s$^{-1}$ \citep[e.g. ][]{Sun2017} to the gas cloud’s current position would therefore have been emitted between 11-14 Myr ago. Given AGN lifecycles exist on much shorter timescales, $10^7-10^8$ yr, while the relic phase after switching off is a further order of magnitude shorter \citep{Parma2007,Sun2017}, it is plausible the outflow originated from a now-passive AGN.\\

Regardless of the velocities, an outflow would need to have a unique geometry with a large opening angle to cover all three arcs (a), (b), and (d). Although, there is evidence of isotropic outflows from highly star-forming galaxies at z$\sim 2$ \cite{Steidel2010} which could make such a geometry possible. It may simply be the case however, that the gas is not from a recent inflow or outflow and therefore it has not retained kinematic signatures of its origin. Recycled gas can take $>1$ Gyr to fall back onto the galaxy \cite{Oppenheimer2012}, and with an estimated 3.3 Gyr since the last significant star formation period it is therefore possible we are detecting previous isotropic outflow that is now reaccreting.\\

Lastly, there is the possibility that the gas detected at these distances (10-20 kpc) is part of an extended outer disk. In the local Universe, the ATLAS 3D survey \citep{Cappellari2011} found 10-40\% of their early-type galaxies contain HI masses above $\sim 10^7 M_{\odot}$ \citep{Serra2012}, with the number fraction varying with environmental density. The HI morphologies range from settled regular disks and rings to irregular distributions suggesting tidal tails and accretion events. However for local early-type galaxies an HI disk always has an ionized gas counterpart \citep{Oosterloo2010,Morganti2006}. Furthermore early-types with inner HI discs are also detected in molecular gas \citep{Oosterloo2010}, and the presence of molecular gas also coincides with ionized gas and $H_{\beta}$ and [OIII] optical emission lines \citep{Crocker2011}. Given J0755 has no detected optical emission lines (the KCWI and GTC/OSIRIS spectra span a continuous rest-frame wavelength of $2030 - 4460${\AA} covering $H_{\beta}$ and the [OII] doublet), it is therefore unlikely to host an HI disk analogous to local early-types. Additionally, given the old stellar age of J0755 (3.3 Gyr) and low current star formation rate of SFR$=0.49^{+0.36}_{-0.38} \rm{M_{\odot} yr^{-1}}$, there has not been any ionized gas for a significant amount of time. Although it is possible that over the next 6.4 Gyr from $z=0.7221$ to now the detected gas may accrete onto the galaxy and form the kind of extended HI disks observed at $z=0$, making J0755 the progenitor of HI rich local early-types.\\

Irrespective of the gas's origins, the question remains whether it will eventually (re)accrete onto the galaxy, and if so, will it restart star formation? Based on the high total mass of the galaxy ($M_E = 10^{12.13} M_\odot$) the velocity offsets of the gas are less then the halo escape velocity, and so the gas is likely bound within the halo \cite{Huang_Yun-Hsin2016} and with enough time should accrete. Considering our estimate of $10^{8.4}-10^{9.3}M_\odot$ of HI gas directly probed within 18 kpc ($\approx 4 R_\mathrm{e}$), this is comparable to the amount of HI gas found distributed in regularly rotating discs within $\sim 20\%$ of early-type galaxies at $z\sim 0$, the presence of which is a strong indicator of on-going star formation \cite{Serra2012}. However, this HI mass measurement is a lower limit on the total gas present in the halo out to higher impact parameters. Estimates of the total cool, photoionised CGM mass around early-type galaxies based on quasar sightlines at $z \sim 0.2$ predict significantly higher ($M_{H} = 10^9 - 10^{11} M_\odot$) amounts of gas \cite{Thom2012}.\\

How then, do quiescent galaxies remain quiescent in the presence of such significant cool gas reservoirs? Previous semi-analytical models \cite{Afruni2019} predict that while cool $10^5 M_\odot$ clouds can exist around massive quiescent galaxies at scales comparable to their virial radii ($\geq 500$ kpc), upon infall these clouds lose 99\% of their mass due to interactions with hot coronal gas. Hence, according to this model, the very inner regions of quiescent galaxy halos ($\leq 20$ kpc) should be devoid of cool gas, thus explaining their continued quiescence. However, in the local Universe between 10-30\% of early-type galaxies show evidence of a more recent secondary star formation period \cite{Donas2007,Schawinski2007}, and while the fraction of stellar mass formed in rejuvenation events is larger for lower mass galaxies, even galaxies with $M_* = 10^{11} M_\odot$ show evidence of secondary star formation \cite{Thomas2010}. Therefore, is our detection of cool gas at $10-20$kpc to this massive quiescent galaxy puzzling, or are we observing this galaxy in the moments before a rejuvenation episode? If we were to watch J0755 over the 6.46 Gyr between $z=0.7224$ and the present day, it is possible we would see a re-ignition of star formation followed by a return to quiescence.\\

This work shows both that the CGM itself is massive enough to affect lensing models, and that we can use the light from galaxy-scale lenses to spatially probe the gas of the foreground quiescent galaxy at close impact parameters. This work provides clear evidence of significant amounts of cool gas at close impact parameters to massive quiescent galaxies, in contrast to the theory that $>$99\% of cool halo gas clouds are destroyed upon infall from the outer halo \cite{Afruni2019}. The survival of these clouds onto the host, and whether they reignite star formation remains a puzzle. Depending upon the exact amount of HI present, a possible consequence of this cool gas is a rejuvenation event which sees J0755 briefly ($\sim 0.7$ Gyr \cite{Chauke2019}) rejoin the star-forming main sequence followed by continued quiescence. This analysis demonstrates the power of galaxy-scale gravitational lens in probing the baryon cycle of massive galaxies.\\

\clearpage

\section*{Methods}

\subsection{KCWI Observations}\label{sec:methods_kcwi_obs}
We observed J0755 with KCWI on the Keck II telescope on 3\textsuperscript{rd} March 2002 UT as part of program ID 2022A\_W230 (PI Barone). We used the medium slicer with the blue low (BL) resolution (R=1800) grating, which provides a resolution of 2.5 {\AA} in the wavelength range 3500–5600 {\AA} in a 16.5" $\times$ 20.4'' field of view. The target was observed in 10 separate 26 minute exposures for a total exposure of 260 minutes. 5 of the 10 exposures used an instrument position angle of $0^\circ$ (North), while the other 5 used a position angle of $90^\circ$ to obtain equal spatial resolution in both spatial directions. We reduced the data using the IDL version of the KCWI data reduction pipeline (DRP) using the standard steps aside from the sky subtraction step which is omitted. We used flux standard star BD33D2642 listed in the KCWI DRP starlist to flux calibrate the data in the last step of the DRP. The data was then further reduced following the procedure outlined by \cite{Nielsen2022} to remove the sky background. \cite{Nielsen2022} found that the standard KCWI DRP (IDL, v1.2.2) leaves a wavelength-dependent gradient at approximate 10\% of the sky level in the data, and therefore developed a custom flat-fielding approach to remedy this issue, which we follow here. The separate exposures were optimally combined using Montage \cite{Jacob2010_Montage}. The spectra are barycentric velocity corrected and are listed in vacuum wavelengths. The full wavelength range for the spectra in the 4 arcs are shown in Supplementary Fig. \ref{fig:full_kcwi_wavelength_range}. Throughout the analysis we assume a flat $\Lambda$ cold dark matter ($\Lambda$CDM) Universe with $\Omega_\Lambda= 0.7$, $\Omega_M= 0.3$, and $H_0 = 70$ km s$^{-1}$ Mpc$^{-1}$, and a Chabrier \cite{Chabrier2003} initial mass function.\\

For each of the 4 knots of the Einstein cross (labelled in Fig. \ref{fig:hubble_kcwi_img}), we spatially integrate the KCWI spectra within 3.8" apertures to create 4 higher S/N spectra. We then fit Voigt profiles to the MgII absorption profiles using the VPfit software \citep[version 12.1;][]{Carswell_Webb_VPFIT2014}, which fits for the gas column density ($\log_{10} N$ atoms cm$^{-2}$), velocity Doppler parameter ($b$ km s$^{-1}$) and redshift ($z$), assuming optically thin gas and a partial covering fraction of unity. From the Voigt profiles we also measure the rest-frame equivalent width $W_r^{\lambda 2796}$. The fits are shown in the panels of Fig. \ref{fig:hubble_kcwi_img} and the fit parameters are summarised in Table \ref{tbl:fits_summary}. For the non-detection in arc (c) we measure a 3$\sigma$ upper limit of $W_{r}^{\lambda 2796} < 0.315$ {\AA} assuming a line width of $b=300$ km s$^{-1}$ based on the S/N and instrumental resolution \cite{Hellsten1998}. We apply the same method to measure a $3\sigma$ upper limit of $W_{r}^{\lambda 2796} < 0.625${\AA} assuming a $b=50$ km s$^{-1}$ line width from an SDSS spectum of an $R_{AB}=19.989 \pm  0.018$ quasar with impact parameter $\rho = 206$ kpc west from the lens system. We note that the strong absorption feature red-ward of the MgII profiles at 4845 {\AA} is interstellar CII${\lambda 1334}$ intrinsic to the arc. \\

In arc (d) we find a statistically significant reduction in flux around the expected location of MgII $\lambda$2796{\AA}, 2803{\AA} as shown by Fig. \ref{fig:hubble_kcwi_img} panel (d), however the widths of the fitted Voigt profiles are unconstrained. This could be due to multiple gas clouds along the line of sight which are unresolved due to the low spectral resolution of KCWI. We obtain an estimate on the column density in arc (d) by constraining the Voigt profile to have $b=310$ km s$^{-1}$.\\

To measure the HI masses, we start by first converting our $W_r^{\lambda 2796}$ measurements to HI column densities following the relations of \cite{Menard_Chelouche2009}. The range in values comes from using either the geometric or arithmetic mean in the $N_{HI}--W_r^{\lambda 2796}$ relation \cite{Menard_Chelouche2009}. The large variation in the relation may reflect two separate sequences, and without knowing which this data is likely to follow we quote both. We then integrate the column densities in the areas probed by the arcs (27.2, 38.7, and 39.2 kpc$^2$ respectively for arcs a, b and d), which are measured from high resolution HST F606W imaging (see Methods subsection Arc Areas and Impact Parameters). The three detections span the closest impact parameters, and we place a $3\sigma$ upper limit on the non-detection in arc (c) which has the largest impact parameter (18.9 kpc).

\noindent\subsection{Stellar properties}\label{sec:stellar_properties}

We used the FAST$++$ (\url{https://github.com/cschreib/fastpp}) software which is a modification of the spectral energy distribution (SED) fitting code FAST \citep{Kriek2009} to measure the stellar mass and star formation rate of the galaxy. Observed HST photometry of the galaxy in the F140W, F814W, and F606W bands as well as its rest-frame optical GTC/OSIRIS spectrum were fitted to constrain the stellar mass of $\log M_* = 10.79 \pm 0.07$ and star formation rate of SFR$=0.49^{+0.36}_{-0.38} \rm{M_{\odot} yr^{-1}}$. We adopted a Chabrier \cite{Chabrier2003} initial mass function  with an exponentially declining star formation history and used a grid of single stellar population models from \cite{Bruzual_Charlot2003}. A uniform screen of dust attenuation for the entire galaxy was assumed following a Calzetti \cite{Calzetti2000} attenuation law. The metallicity was set as a free parameter varying between $Z = 0.004$ (sub-solar) and $Z = 0.05$ (super-solar). While more ground-based photometry is available (\textit{ugriz} from DECaLS and SDSS) the deflector light suffers severe contamination from the source light in all the ground-based photometry due to the large point-spread function. We therefore only included the high resolution HST images to ensure clean deflector photometry.\\

To measure the average stellar population age and stellar velocity dispersion of J0755 we use archival rest-frame optical slit spectroscopy from OSIRIS on the Gran Telescopio Canarias. Specifically we used OB0012 of program ID GTC47-16B \cite{Marques-Chaves2020}, for which the slit was oriented across arcs (d) and (b). The 60 minute total observation used the R2500R grating, and spanned the observed wavelength range $\lambda 5575 - 7685$. We reduced the data using the \texttt{PypeIt} \citep{pypeit:joss_pub, pypeit:zenodo} (\url{https://pypeit.readthedocs.io/en/latest/}) Python package.\\

We measure the stellar age of J0755 by fitting synthetic spectral templates from the Extended Medium resolution INT Library of Empirical Spectra \citep[E-MILES][]{Sanchez-Blazquez2006,Vazdekis2016} using the Python implementation of the Penalized Pixel-Fitting software \citep[\texttt{pPXF}][]{Cappellari_Emsellem2004,Cappellari2017}. The analysis used  a $12^{th}$ order multiplicative polynomial to account for the shape of the continuum (relative to the templates) and for dust extinction (see e.g. \cite{Barone2020}). The template library is based on isochrones from \cite{Girardi2000} and a Chabrier \cite{Chabrier2003} initial mass function. We use the full set of 350 templates which contain 50 age values ranging from 0.063 to 17.78 Gyr, 7 [Z/H] values ranging from $-2.32$ to $+ 0.22$, and have [$\alpha$/Fe] values scaled to the solar neighbourhood \citep[the `base’ models][]{Vazdekis2016}. The bestfit model has an age of 3.3 Gyr and stellar metallicity of [Z/H] = -0.058 and is shown in Supplementary Fig. \ref{fig:deflector_OSIRIS}. The stellar kinematics are similarly measured using \texttt{pPXF}, however this time using the MILES stellar library \cite{Sanchez-Blazquez2006} which consists of 985 stars spanning a large range in stellar atmospheric parameters, as well as a $12^{th}$ order additive polynomial. The bestfit model has a velocity dispersion of $304 \pm 13$ km s$^{-1}$.\\

\noindent\subsection{Environment}\label{sec:environment_discussion}

We briefly consider whether close neighbours could be contributing to the MgII absorption detected in the KCWI data as well as the complexity in the lens model. Fig. \ref{fig:hubble_kcwi_img} shows a 2''x2'' view around J0755 from DECaLS \textit{grz} imaging. J0755 is isolated or at most in a sparsely populated group, with no close neighbours that are likely at the same redshift. Both the standard PL model and the multipole PL lens model include the contribution from the $z=0.283$ foreground quiescent galaxy to the south-west, however including or excluding this foreground galaxy does not result in an appreciable change in either model. Furthermore, we do not expect the broader environment to create the type of residuals shown in the top row of Fig. \ref{fig:lens_model}. The gravitational effects of neighbours and the broader environment manifests as an external shear which is already accounted for in the standard PL model. We therefore conclude our results are unlikely to be due to other galaxies in the local environment.\\

\noindent\subsection{Virial Radius}\label{sec:virial_radius_measurement}

To estimate the virial radius of J0755 we first measure the virial mass using the stellar to halo mass relation \cite{Girelli2020}. The corresponding virial radius is then determined using the relation $R_{\rm vir}=[3M_{\rm h}/(200 \rho_c 4 \pi)]1/3$ where $\rho_c$ is the critical density of the Universe at z=0.7224.\\

\noindent\subsection{Arc Areas and Impact parameters}\label{sec:impact_parameter_measurements}

To measure the size of the arcs and their projected distance to the deflector we use the Source Extractor \cite{Bertin1996} software on the HST F606W photometry. First we measure the total flux of the 4 arcs within 1$\sigma$ of the background noise. We then adjusted the $\sigma$ limits until the size contained 90\% of the total light within each arc and which better represents the light included in the KCWI spectra. We then measured the projected distance from the area-weighted mean of the 95th percentile isophotes of the 4 arcs a, b, c and d from the deflector to be 14.4, 14.6, 18.9, 11.1 kpc respectively. We measure the sizes of the arcs to be 0.52, 0.74, 0.26, and 0.75 sq" which, based on the assumed cosmology, translate to 27, 39, 14 and 39 kpc${^2}$ (proper), giving a total area probed by the arcs of 119 kpc${^2}$.\\

To compare the area probed by the arcs measured above to the area probed by a typical quasar, we assume the following. Firstly that the the scale of the quasar probe derives from broad line region, which are on the order of $\sim 40$ lightdays \cite[see Fig 6. of ][]{Mandal2021}, which translates to 0.03 pc. Therefore the area probed by a quasar is $\pi 0.03^2 = 0.004$ pc$^{-2}$. Hence based on these measurements, the gravitational arcs of J0755 probe a spatial area 10$^{11}$ greater than a quasar sightline.\\

\noindent\subsection{Lens model}\label{sec:hst_imaging_and_lens_model}

To construct a lens model we used publicly available HST/WFC3 UVIS F606W data from program 14189 (PI: Bolton). We perform lens modeling using the open-source software {\tt PyAutoLens} (\url{https://github.com/Jammy2211/PyAutoLens}) \cite{Nightingale2015,Nightingale2018,Nightingale2021} using an adaptation of the Source, Light and Mass (SLaM) pipelines described in \cite{Etherington2023} and \cite{Nightingale2023}. The specific pipelines used in this study are available at \url{https://github.com/Jammy2211/autolens_j0755}. Lens models are fitted using the nested sampling algorithm {\tt nautilus} \cite{Lange2023}. We give a concise overview of the aspects of lens modeling that are most important for this study, however a detailed step-by-step guide of the lens analysis is provided at \url{https://github.com/Jammy2211/autolens_likelihood_function}.\\

To subtract the deflector galaxy's light, we perform a multiple Gaussian expansion (MGE) \cite{Cappellari2002}, which decomposes the lens light into 60 elliptical two dimensional Gaussians. The axis-ratio, position angles and size of the Gaussians vary, capturing departures from axisymmetry. The intensity of every Gaussian is solved for simultaneously with the source reconstruction, using a non-negative least square solver (NNLS). Previous analyses assumed S\'ersic profiles for the lens subtraction (e.g. \cite{Nightingale2022}) and obtained the same conclusions regarding there being source residuals after the model is fitted, indicating the lens light subtraction is not important for our interpretation of an additional diffuse mass component in the lensing galaxy.\\

The aspects of lens modeling that are important for this study are the lens galaxy mass model (which ray-traces light to the unlensed source-plane) and the source analysis (which for a given mass model reconstructs the source's unlensed light). The source is reconstructed using an adaptive Voronoi mesh with 2000 pixels whose distribution adapt to the source morphology. This uses the natural neighbour interpolation and adaptive regularization described in \cite{Nightingale2023}, but unlike this study enforces positivity on source pixels by using the NNLS. This Voronoi mesh is able to reconstruct irregular galaxy morphologies, provided the mass model is accurate. A user input mask defines where the source reconstruction is performed. The mask used in this study is shown in Fig. \ref{fig:lens_model} (black line). It has been customized via a graphical user interface to only contain the brightest arcs which the KCWI specta indicate have CGM absorption. It also contains a second faint additional lensed source, where one multiple image is located to the top-right of the masked image.\\

We present two models in the main paper. The first assumes a power-law (PL) mass model \citep{Tessore2015} with an external shear term, representing the lens galaxy's total mass distribution (stars and dark matter). The second assumes a power-law plus external shear with internal multipoles \citep{Chu2013}, which describe internal angular structure in the lens mass distribution. Different multipoles of different order $m$ correspond to different forms of complexity. For example, the $m=1$ multipole accounts for offsets in the lens's centre of mass and an $m=4$ accounts for boxiness or diskiness. Specifically, we fit simultaneously a PL plus shear with an $m=1$, $m=2$, $m=3$ and $m=4$ multipole, with the rationale that in unison they may be able to capture the asymmetric and diffuse mass structure of the CGM. Equations for all mass models are given at \url{https://github.com/Jammy2211/autolens_j0755}.\\

To the south-west of J0755 is a foreground passive galaxy, which can be observed in all HST images. CaII K and H ($\lambda \lambda 3934, 3969$) detected in the KCWI spectra reveals this galaxy is at $z=0.283$. We include it in the lens model as an Singular Isothermal Sphere mass distribution, with the centre fixed to the observed centre of light. Multi-plane ray tracing is performed to account for the galaxy being at a different redshift to the lens. A second fainter galaxy to the north-west of J0755 is also included in the lens model.\\

Including these galaxies, the PL plus shear mass model has 10 free parameters compared to the PL with multipoles plus shear which has 18. To objectively compare two models with such different degrees of freedom we use the Bayesian evidence, $\mathcal{Z}$, which is estimated by {\tt nautilus}. The evidence is the integral of the likelihood over the prior and therefore naturally includes a penalty term for including too much complexity in a model. We quote Bayes factors defined as $\Delta\,\mathrm{ln}\,\mathcal{Z} = \mathrm{ln}\,\mathcal{Z}_{\rm PLM} - \mathrm{ln}\,\mathcal{Z}_{\rm PL}$, where $\mathrm{ln}\,\mathcal{Z}_{\rm PLM}$ corresponds to the PL with multipoles plus shear and $\mathrm{ln}\,\mathcal{Z}_{\rm PL}$ the PL plus shear. A value of $\Delta\,\mathrm{ln}\,\mathcal{Z} = 4.5$ corresponds to odds of 90:1 in favour of that model; a $\rm 3\sigma$ preference. A value of $\Delta \ln \mathcal{Z} = 12.5$ corresponds to a $\rm 5\sigma$ preference.\\

The main results of this analysis can be summarised as: (i) Bayesian model comparison favours the PL with multipoles model with a Bayes factor of $\Delta\,\mathrm{ln}\,\mathcal{Z} = 574 (67$) for the F606W (F814W) data (both corresponding to an over $10\sigma$ preference); (ii) the models for both filters favour a first order multipole component with a magnitude over $10\%$ indicating a mass component with a significant offset from the stars and dark matter and; (iii) the azimuthal structure introduced by the multipoles creates the asymmetric squeezing seen in the convergence plots of Fig. \ref{fig:lens_model}. Light from J0755's lens galaxy also shows no evidence of asymmetry making it unlikely this mass is present in stars or dark matter.\\

We consider whether we can extract a mass estimate of only the CGM component via the lens model. For example, we inspected the convergence residuals in panel (g) of Fig. \ref{fig:lens_model}, which suggest the component has a mass above $10^{10} M_\odot$. However, the Einstein mass of the PL with multipoles model is less than $95\%$ that of the PL model (systematic errors on Einstein masses above $5\%$ are expected, e.g. \cite{Etherington2023}). The magnitude of these residuals is therefore driven by the overall convergence normalization reducing by $5\%$. Given an Einstein mass of $\log M_{\rm E}/M_\odot = 12.13$, this produces changes in convergence mass above $10^{10} M_\odot$. Extracting a mass estimate of the CGM from the multipole model also proved unsuccessful. This is because the multipole model does not add a separate mass component on top of the PL, but redistributes its mass, such that the overall mass associated with the multipoles alone averages to zero. This means it is not possible to provide a precise estimate of the mass associated with only the off-centre and elliptically squeezed component.\\

\newpage

\section{Acknowledgements}

This research was supported by the Australian Research Council Centre of Excellence for All Sky Astrophysics in 3 Dimensions (ASTRO 3D) through project CE170100013. TMB, KG, and TN acknowledge support from Australian Research Council Laureate Fellowship FL180100060. TMB and KG acknowledge support from Australian Research Council Discovery Project DP230101775. JvdS acknowledges the support of an Australian Research Council Discovery Early Career Research Award (project number DE200100461) funded by the Australian Government. S.L. acknowledges support by FONDECYT grant 1231187. Some of the data presented herein were obtained at the W. M. Keck Observatory, which is operated as a scientific partnership among the California Institute of Technology, the University of California and the National Aeronautics and Space Administration. The Observatory was made possible by the generous financial support of the W. M. Keck Foundation. Observations were supported by Swinburne Keck program with KCWI 2022A\_W230. The authors wish to recognize and acknowledge the very significant cultural role and reverence that the summit of Maunakea has always had within the Native Hawaiian community. We are most fortunate to have the opportunity to conduct observations from this mountain.

Software: Astropy \cite{astropy:2022}, Matplotlib \cite{Hunter:2007}, NumPy \cite{harris2020array}, SciPy \cite{2020SciPy-NMeth}, QFitsView \cite{Ott2012_QFitsView}, and SAOImage DS9 \cite{SAOImage_DS9_2000}.

\section{Competing Interests}
The authors declare no competing interests.

\section{Author contributions}
TMB led the project including directing the observations, measuring the equivalent widths, performing the stellar population modelling, and writing the paper. GGK contributed significantly to the direction of the analysis and interpretation of the results. JWN performed the lens modelling and contributed significantly to their interpretation. NMN reduced the KCWI spectra and contributed to the interpretation of results. HN performed the SED fitting to measure the stellar mass. HS contributed to the reduction of the Hubble images. KG, KVHT, TJ, TN, and SL all contributed to the interpretation. KVGC, NS and GL contributed to the initial sample as part of the ASTRO3D Galaxy Evolution with Lenses (AGEL; \cite{Tran2022}) project. JvdS contributed to the comparison with ATLAS3D survey. All coauthors commented on this manuscript as part of an internal review process.

\section{Data Availability}
Raw Keck/KCWI data are publicly available at the Keck Observatory Archive (https://www2.keck.hawaii.edu/koa/public/koa.php) under program ID W230. Hubble Space Telescope imaging is publicly available on the Barbara A. Mikulski Archive for Space Telescopes under program IDs GO-14189, SNAP-15867, and GO-16734. Raw GTC/OSIRIS spectra are publically available from the Gran Telescopio Canarias Public Archive (https://gtc.sdc.cab.inta-csic.es/gtc/index.jsp) under program ID GTC47-16B.

\begin{suppfigure*}
    \includegraphics[width=\textwidth]{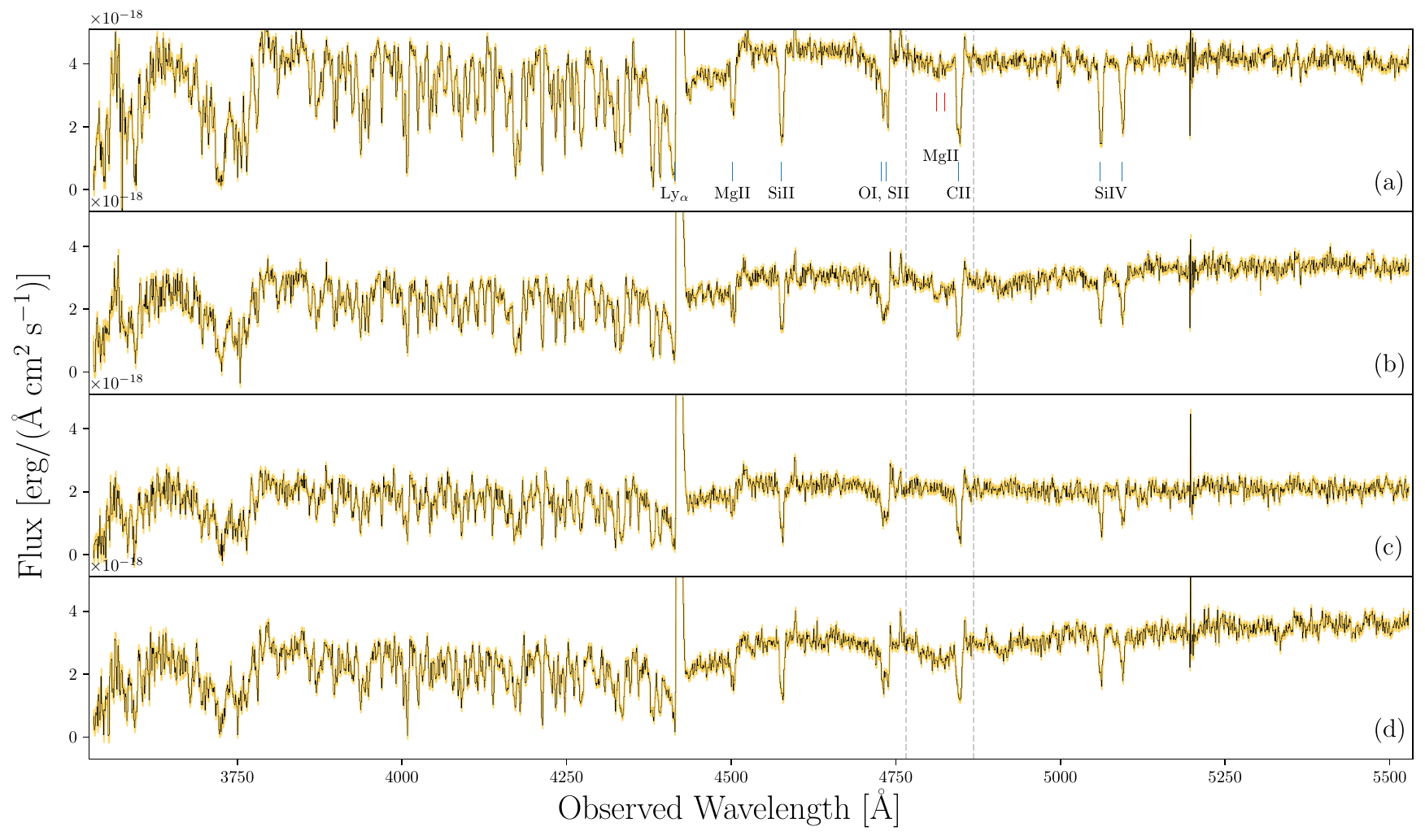}
    \caption{\textbf{Integrated Keck/KCWI optical spectra of the 4 arcs labelled a-d.} The black line is the observed spectrum, the yellow shaded region is the $1\sigma$ uncertainty. The grey dashed vertical line indicate the regions shown in Figure \ref{fig:hubble_kcwi_img}. The blue ticks indicate interstellar medium emission and absorption lines from the z=2.6347 source. The red ticks show the MgII $\lambda \lambda 2796, 2803$ from the CGM of the z=0.7224 deflector.}\label{fig:full_kcwi_wavelength_range}
\end{suppfigure*}

\begin{suppfigure*}
    \centering
    \includegraphics[width=\textwidth]{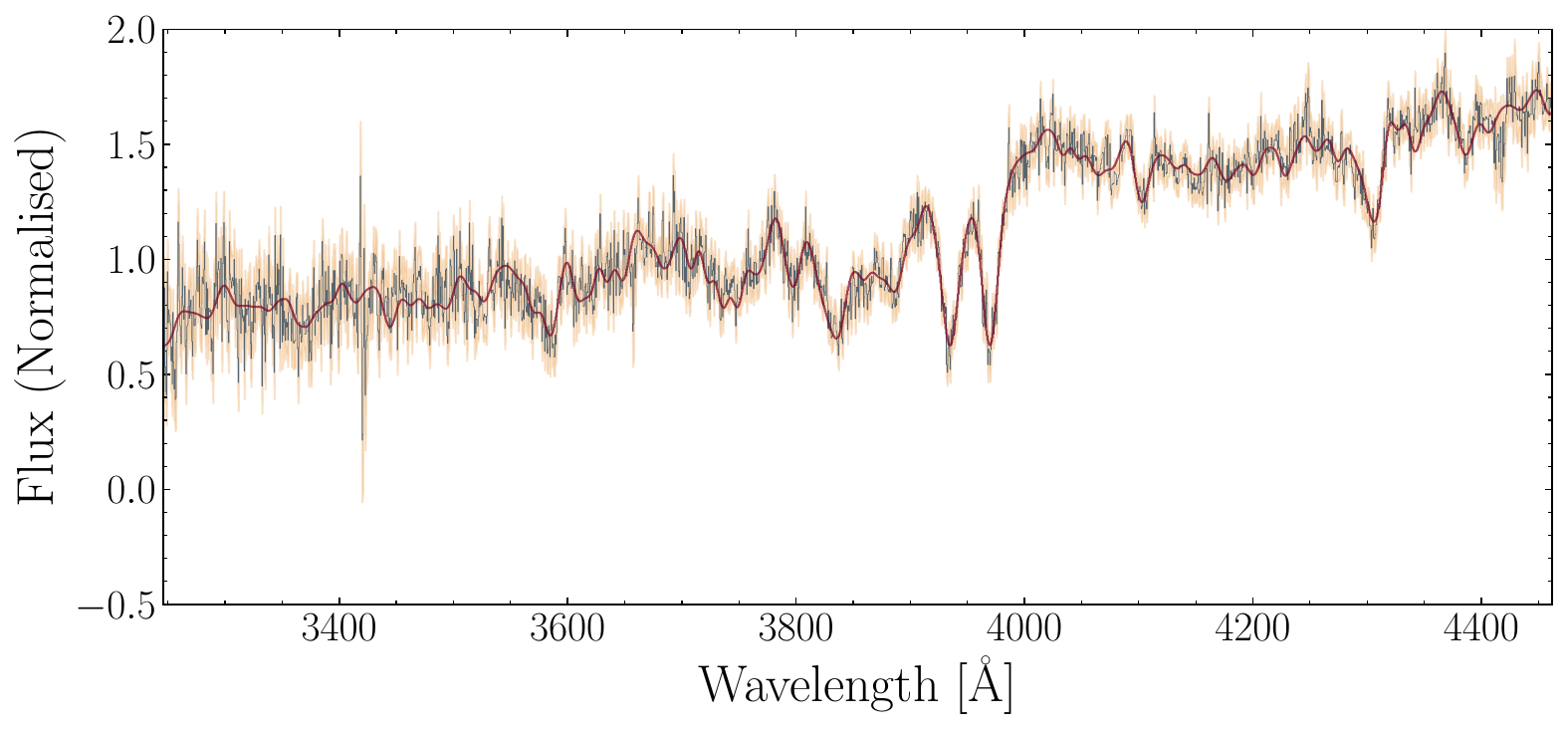}
    \caption{\textbf{Rest-frame optical spectrum of the massive quiescent deflector taken by the OSIRIS slit spectrograph on the Gran Telescopio Canarias}. The grey line is the observed spectrum, the orange shaded region is the 1$\sigma$ uncertainty, and the red line is the best-fit from the stellar populations model.}
    \label{fig:deflector_OSIRIS}
\end{suppfigure*}

\begin{suppfigure*}
    \centering
    \includegraphics[width=\textwidth]{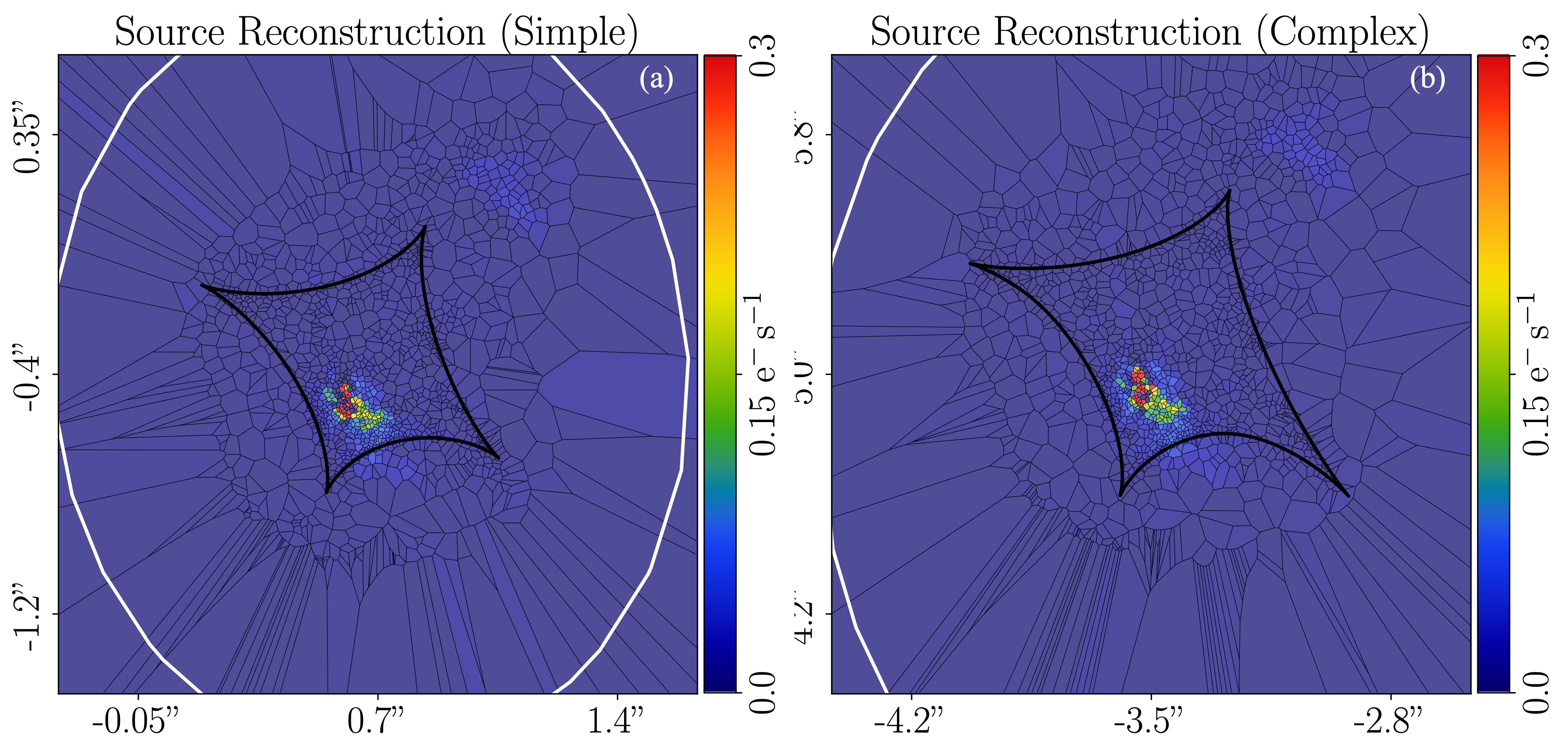}
    \caption{\textbf{Voronoi Source Reconstructions for the two lens models}. \textit{(a)}: The simple power law plus external shear lens model reconstruction. \textit{(b)}: The power law plus multipoles model. The solid black line shows the tangential caustic and the white line is the radial caustic derived from the lens model.}
    \label{fig:source_reconstruction}
\end{suppfigure*}

\begin{suppfigure*}
    \centering
    \includegraphics[width=\textwidth]{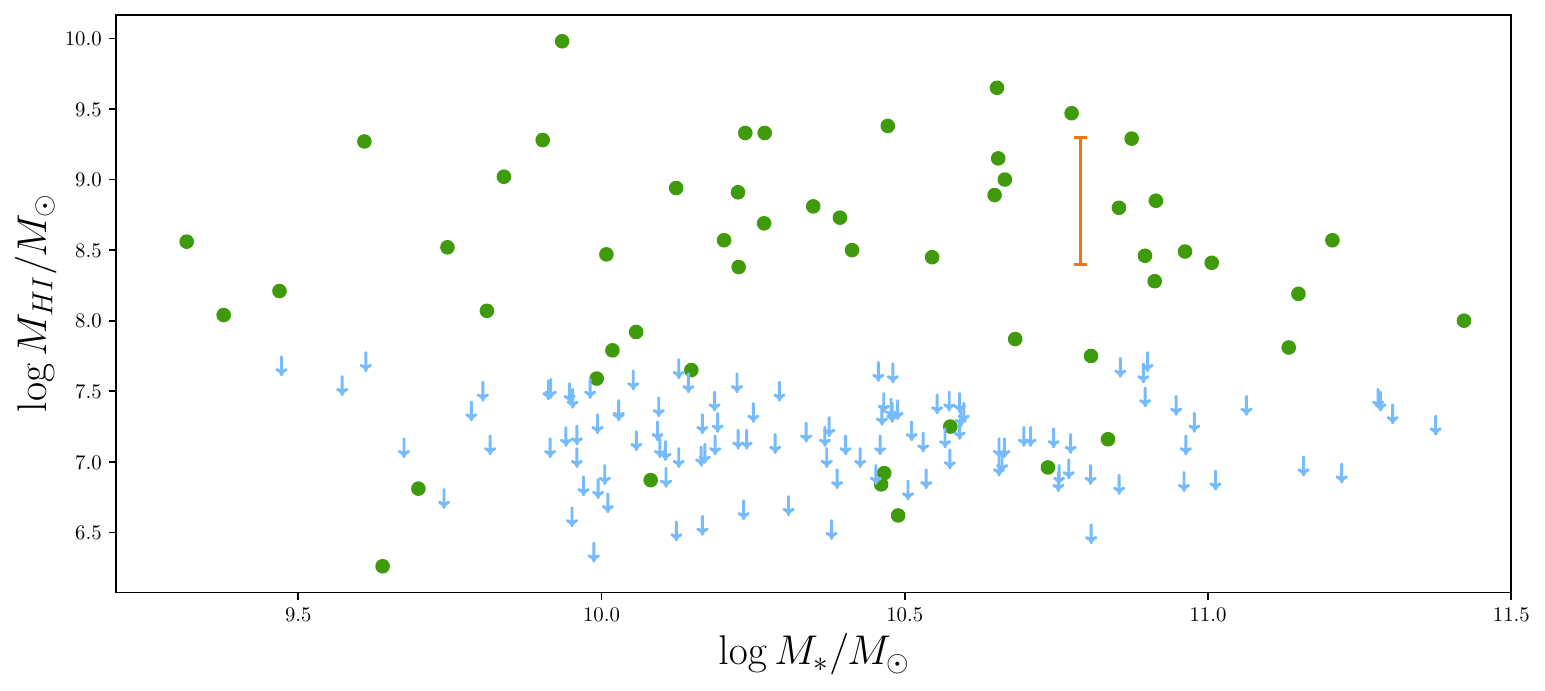}
    \caption{\textbf{Comparison between the stellar mass and HI mass for local ETGs from the ATLAS3D survey \cite{Serra2012,van_de_Sande2019} and J0755}. Green points are ATLAS3D HI detections, blue arrows are upper limits on non-detections. The orange point is the measured HI mass within 18kpc ($10^{8.4} - 10^{9.3} M_{\odot}$) for J0755. We see that the J0755 measurement is consistent with the amount of HI gas detected in local ETGs.}
    \label{fig:atlas3d_HI_gas_mass}
\end{suppfigure*}

\begin{suppfigure*}
    \includegraphics[width=\textwidth]{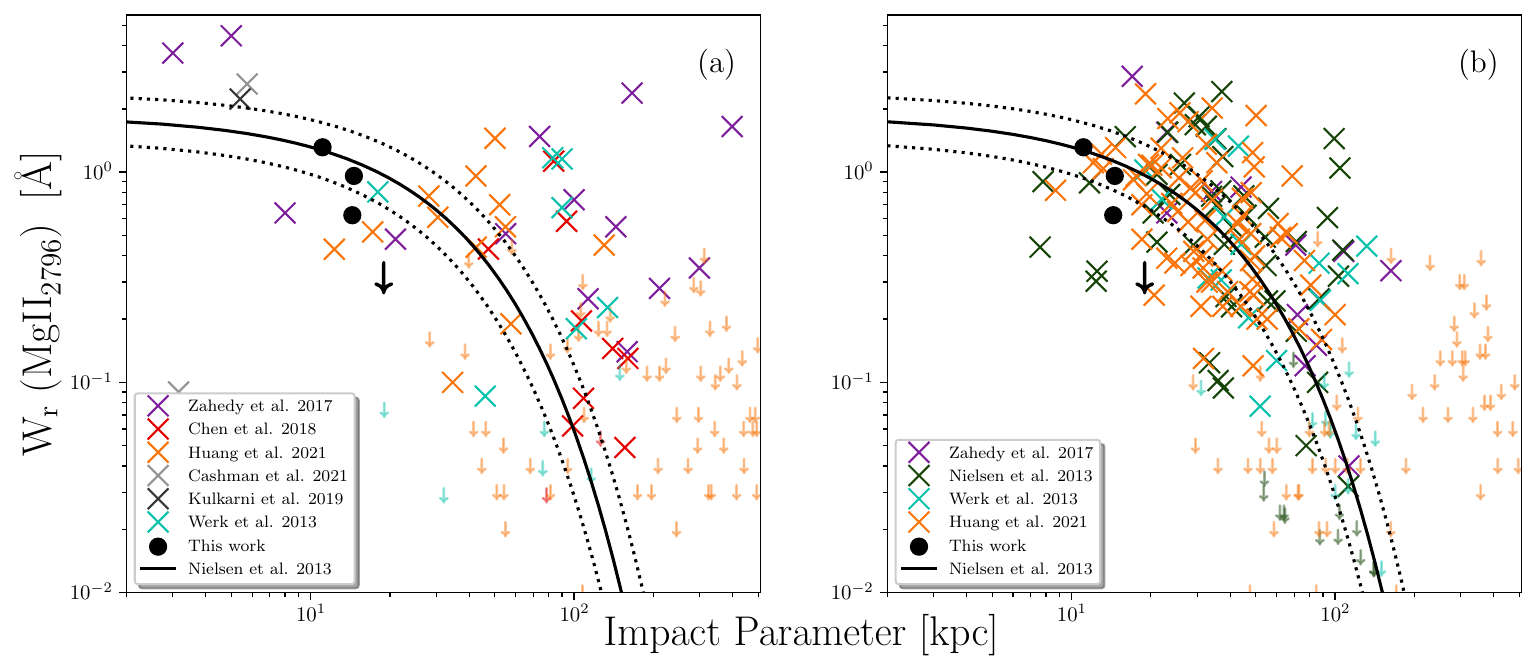}
    \caption{\textbf{Comparison of the equivalent width of MgII versus impact parameter of J0755 to galaxies probed via quasar sightlines.} \textit{(a)}: Quiescent galaxies, and \textit{(b)}: Star-forming galaxies. Definition of quiescent and star-forming taken from the literature studies. Both panels show the relation expected for star-forming galaxies from \cite{Nielsen2013}.}\label{fig:quasar_literature_comparison}
\end{suppfigure*}

\clearpage

\bibliography{cgm_bibliography}

\end{document}